\documentclass[usegraphicx,useAMS,usenatbib]{mn2e}
\newcommand{\be}{\begin{equation}}
\newcommand{\ee}{\end{equation}}
\newcommand{\bd}{\begin{displaymath}}
\newcommand{\ed}{\end{displaymath}}

\usepackage{graphicx}
\usepackage{longtable} 
\usepackage{longtable} 
\usepackage{rotating}

\usepackage{times}

\title[Starspot size on young active stars]
 {On the relationship between the size and surface coverage of starspots on magnetically active low-mass stars.}

\author[R. J. Jackson and R. D. Jeffries]
  {R. J.~Jackson and R. D.~Jeffries\\
   $^1$ Astrophysics Group, Research Institute for the Environment, Physical Sciences and Applied Mathematics, Keele University,\\
   $\;$ Keele, Staffordshire ST5 5BG, UK}

\date{Accepted for publicaton in MNRAS}

\pagerange{\pageref{firstpage}--\pageref{lastpage}} \pubyear{2012}

\def\LaTeX{L\kern-.36em\raise.3ex\hbox{a}\kern-.15em
    T\kern-.1667em\lower.7ex\hbox{E}\kern-.125emX}

\begin{document}
\label{firstpage}
\maketitle

\begin{abstract}

We present a model that predicts the light curve amplitude distribution
for an ensemble of low-mass magnetically active stars, under the
assumptions that stellar spin axes are randomly orientated and that
cool starspots have a characteristic scale length and are
randomly distributed across the stellar surfaces. The model is compared
with observational data for highly magnetically active M-dwarfs in the
young cluster NGC~2516. We find that the best fitting starspot scale
length is not constrained by these data alone, but requires assumptions
about the overall starspot filling factor and starspot temperature.
Assuming a spot coverage fraction of $0.4\pm 0.1$ and a starspot to
unspotted photosphere temperature ratio of $0.7\pm 0.05$, as suggested
by the inflated radii of these stars compared to evolutionary model
predictions and by TiO band measurements on other active cool stars of
earlier spectral type, the best-fitting starspot angular scale length
is $3.5^{+2}_{-1}$ degrees, or a linear scale length of $\sim
25\,000$\,km.   This linear scale length is similar to large sunspot groups, but 2--5 times
smaller than the starspots recently deduced on an active G-dwarf using
eclipse mapping by a transiting exoplanet. However, the best-fitting
spot scale length in the NGC 2516 M-dwarfs increases with the assumed
spot temperature ratio and with the inverse square root of the assumed
spot filling factor. Hence the light curve amplitude distribution might
equally well be described by these larger spot scale lengths if the
spot filling factors are $< 0.1$ or the spot temperature ratio is $>
0.9$.

\end{abstract}

\section{Introduction}

Starspots are a ubiquitous manifestation of magnetic
activity in the photospheres of cool stars with convective
envelopes. Their sizes, filling factors and temperatures are important
constraints on the dynamo mechanism, which regenerates and amplifies
the sub-photospheric magnetic field, and on the magneto-hydrodynamic 
processes which shape
the emergence of magnetic fields from these sub-photospheric layers out into
the photosphere and beyond (see reviews by Thomas \& Weiss 2008;
Strassmeier 2009).
Beyond these diagnostic roles, starspots cause
rotational modulation of light curves that enable stellar rotation
periods to be estimated,  are a nuisance source of radial velocity
jitter when searching for exoplanets (e.g. Reiners et al. 2010; Barnes,
Jeffers \& Jones 2011), confuse the estimation of stellar radii in active,
eclipsing binaries (e.g. Jeffers et al. 2006; Morales et al. 2010) 
and, if the filling factor is large, could significantly alter the
structure of low-mass stars by blocking convective flux in their outer
envelopes, leading to increased radii and decreased effective
temperatures (e.g. Spruit \& Weiss 1986;  Chabrier, Gallardo \& Baraffe
2007; MacDonald \& Mullan 2012).

Jackson, Jeffries \& Maxted (2009) estimated the average radii of
fast-rotating late K- and M-dwarfs in the young open cluster NGC~2516,
by multiplying together their rotation periods and equatorial
velocities. These magnetically active stars appear to have radii that
are larger than both model predictions and the radii measured by
interferometry for otherwise similar, magnetically inactive
field stars. The radius discrepancy increases towards lower masses,
reaching $\simeq 50$ per cent at a given luminosity in M4
stars. Jackson et al. interpreted this inflation in terms of a
two-temperature photospheric model that required large (20 per cent to more than
50 per cent in the coolest stars) filling factors of dark starspots.

The dark starspot hypothesis was motivated by: (i) the qualitatively
similar discrepancies seen in the magnetically active components of
close, low-mass binary systems (Ribas at al. 2008; Morales et al. 2009;
Torres, Andersen \& Gimenez 2010), for which a similar explanation has
been advanced (Chabrier et al. 2007; 
Morales et al. 2010); (ii) the
similarity of the proposed spot filling factors and temperatures to
those determined for very active G- and K-stars from careful modelling of
their optical TiO absorption bands (filling factors of 20--50 per cent and
temperature ratios between spots and unspotted photosphere of 0.65--0.76; O'Neal,
Neff \& Saar 1998; O'Neal et al. 2004; O'Neal 2006).

In addition to yielding rotation periods, the broadband light curves of
magnetically active stars contain information about the distribution of
spots on the stellar surface. 
Jackson et al. (2009) used rotation periods determined from I-band
light curves that had typical first harmonic amplitudes of only
0.01-0.02 mag (from Irwin et al. 2007). Furthermore, Jackson \&
Jeffries (2012) showed that about half of the monitored low-mass
members of NGC 2516, from the same data set, had no detectable light
curve modulation at all, despite being just as magnetically active
(judged by their chromospheric emission) as their periodic siblings and
having a similar distribution of equatorial rotation velocities.  It
seems paradoxical to propose that such stars have large spot filling
factors yet such small light curve amplitudes, but Jackson \&
Jeffries (2012) pointed out that the detectability of any rotational
modulation might simply be governed by the degree of axisymmetry of the
starspot distribution. The paradox might be resolved if the large
filling factors were made up of many, randomly placed dark spots with
typical angular diameters of $\sim 2$ degrees on the surface.

In this paper we place the scenario described by Jackson \&
Jeffries (2012) on a quantitative basis by presenting a simple
numerical starspot model that predicts the light curve properties of
an ensemble of active stars. We compare these model predictions with
the observed properties of low-mass  (M0--M4) stars in NGC 2516 and explore the
relationship between spot filling factor, spot temperature, spot scale length
and the distribution of light curve amplitudes. 
In section 2 we describe our model and its key assumptions; section
3 presents the model results and how well the model parameters are
constrained by the observations; section 4 discusses the results in the
context of the solar-stellar analogy and efforts to determine spot
parameters using other techniques.

\begin{figure}
\centering
	\includegraphics[width = 84mm]{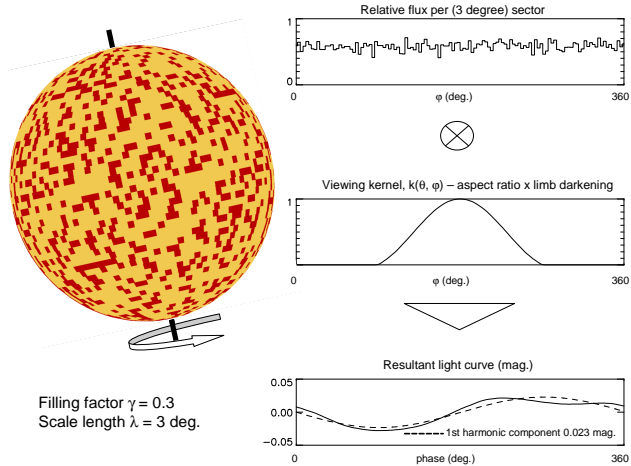}
	\caption{Simple model of a spotted star with a random covering
          of starspots of uniform area used to predict the probability
          density of light curve modulation amplitudes. The right hand
          panels illustrates how the flux from randomly distributed spots
          adds to produce the stellar light curve.
}
	\label{fig1}
\end{figure}

\section{Calculation of light curve modulation amplitudes}

 In low-mass stars with a central radiative zone it is likely
  that an ``$\alpha \Omega$'' dynamo is responsibly for amplifying
  magnetic field at the boundary between the radiative zone and
  convective envelope (Parker 1975). On the Sun this gives rise to
  latitude-dependent spot coverage and a strong latitude dependence has
  been predicted in fast-rotating stars with deep convective envelopes
  (such as the late-K to M4 dwarfs considered here), such that
  starspots will emerge predominantly at high latitudes (Sch\"ussler \&
  Solanki 1992). In some cases this is confirmed by Doppler tomography of
  young, fast rotating K-stars which can have polar spots, but also some
  spots at lower latitudes (e.g. Stout-Batalha \& Vogt 1999; Jeffers,
  Donati \& Collier Cameron 2007). However, in other cases the
    spots on K-stars seem
  evenly distributed at 
  all latitudes (Barnes et al. 2001). In early M-dwarfs there is no
  evidence for any strong latitudinal dependence of spot position from
  Doppler images (Barnes \& Collier Cameron 2001). Some of
  our sample have spectral types cooler than M3.5, at which point the
  radiative core disappears and the nature of the dynamo may change to
  a turbulent ``$\alpha^2$'' dynamo (Chabrier \& K\"uker 2006). 
  There is little observational information on how spots might be
  distributed on the surface as a result. In what follows we will adopt the
  simplest assumption -- that spots are randomly distributed on the
  surfaces of all our sample stars.

For simplicity, we also assume that any periodic light curve variations
are due to dark starpots. In comparatively
low activity stars like the Sun there is also a contribution from
bright plages or faculae. Comparison of chromospheric and photospheric
activity, suggests that the contribution of plages and faculae diminishes
in more active stars (Radick et al. 1998; Lockwood et al. 2007).

Figure 1 illustrates the model used to predict the effects on the light
curve amplitude produced by a random distribution of small
starspots on the stellar surface. The surface of the star is divided
into nominally equal cells of  solid angle $\lambda^2$, where $\lambda$
is an angular scale length. Surface luminosities are randomly assigned to
these cells according to the average starspot filling factor, $\gamma$
and spot luminosity ratio, $\kappa$. When calculated for a set of
stars, the resultant distribution of
light curve amplitudes depends on four parameters:

\begin{itemize}
	\item \textbf{Scale length}, $\lambda$ is the angular distance
          between areas on the stellar surface that can show
          independent starspot activity. The absolute linear scale
          lengths for stars of different
          radii, $R$, can be compared by considering
          an angular size $\lambda R/R_{\odot}$, i.e. the equivalent
          angular distance on the surface of the sun. $\lambda
          =2$ degrees corresponds to a cell covering 0.01 percent of the
          stellar surface. $\lambda$ is not quite the same as the mean
          starspot size since the latter depends on how randomly
          distributed areas of starspot activity group together on the
          stellar surface which in turn depends on filling factor (see
          the example of a spotted star in Fig. 1).

  \item \textbf{Filling factor, $\gamma$} is the fractional area
  covered by starspots. A fraction (1-$\gamma$)
  of the stellar surface is unaffected by starspot activity, with a
  surface flux corresponding to the effective temperature, $T_o$, of an
  unspotted star. The remaining fraction $\gamma$ shows a reduced
  surface flux depending on the spot temperature. This spot temperature
  need not be uniform; spots may comprise an umbra, with a large 
  temperature reduction, surrounded by a penumbra at intermediate
  temperatures.

  \item \textbf{Luminosity ratio, $\kappa$} is the ratio of the average
    surface flux of a spotted cell to that of an unspotted cell in the
    wavelength band of the measured light curve.  In the simple case of
    a spot with a uniform temperature $T_s$, the $\kappa$ value for the
    $I$-band light curves considered here declines as $(T_o/T_s)^n$,
    where $n \sim 5$ due to the usual Stefan's law combined with the
    temperature dependence of the $I$-band bolometric correction. It
    turns out that $\kappa$ cannot be constrained by the data we model
    and must therefore be assumed. The work of O'Neal et al. (1998),
    O'Neal et al. (2004) and O'Neal (2006) suggests that $T_s/T_o$ lies
    in the range 0.65 to 0.76 for very active G- and K-dwarfs (that are
    somewhat warmer than our sample). Light curve modelling of M-dwarfs
    (e.g. Berdyugina 2005; Rockenfeller, Bailer-Jones \& Mundt 2006) suggests that
    spots may only be a few hundred Kelvin cooler than the unspotted
    photosphere, but these crude, single-spot models will greatly
    overestimate the spot temperature if there are many smaller
    spots. We consider a range of possiblilities from
    $0.5<T_s/T_o<0.9$, corresponding to $0.03<\kappa<0.59$ for $I$-band
    light curves.
  
 \item \textbf{Completeness scale, $\sigma$}. The
 proportion of targets that will yield a {\it measured} rotation period
 as a function of light curve amplitude is characterised 
  as a one-sided cumulative Gaussian distribution with a
 standard deviation of $\sigma$ (in magnitudes), 
 i.e. the completeness function varies
 from zero for small light curve amplitudes to unity for large light
 curve amplitudes.  The value of $\sigma$~is a free parameter in
   our model that depends on the
 observation cadence and sensitivity, which we will assume are uniform
 for a particular survey. However, within a survey, $\sigma$ is likely
 to vary with target star brightness.
  
\end{itemize}

\subsection{Calculation procedure}
A Monte Carlo method is used to model the effects of randomly
  placed star spots on the the probability distribution of light curve
  amplitudes for a grid of scale lengths, filling factors and
  luminosity ratios. For each combination;
\begin{itemize}
	\item The surface of the star is divided into cells of
          nominally equal solid angle of $\lambda^2$ steradians  that are arranged
					in strips of constant latitude.  A fraction
          $\gamma$ of these cells are assigned a surface flux density of
          $\kappa$ relative to the flux density of the unspotted surface. The
          selection of which cells are darker is made randomly.

	\item   The flux densities from individual cells in each latitudinal strip are
          re-binned into longitudinal bins matching the longitudinal size of the original 
          cells at the equator to give equal numbers of cells at all latitudes.        

	\item   The contribution from each of these re-binned cells are
          summed according to their area, viewing angle and limb
          darkening to give the stellar luminosity as a function of
          rotation phase.
	
	\item The variation of luminosity (relative to the mean) is analysed to determine 
	the magnitude of the first harmonic component of light curve amplitude. 

	\item Results of repeated simulations are accumulated to determine the probability 
	distribution of light curve amplitudes as a function of $\lambda$, $\gamma$  and $\kappa$.

\end{itemize}

To calculate the light curve the net flux density of each of the rebinned cells 
 is first scaled according to the cell area (which varies as
$\cos \theta$ where $\theta$ is the cell latitude) to give the net flux per cell.
Fluxes from the cells in each latitudinal strip are then convolved 
with a viewing kernel,$k$ and the result summed over all latitudes to give the 
light curve of the spotted star as a function of rotation,  where; 
\begin{equation}
k = \cos \theta \cos \phi(1-\mu(1-\cos \theta \cos \phi)) \;\; 
{\rm for}\, \frac{-\pi}{2} < \phi < \frac{\pi}{2}
\end{equation}

\noindent{where $\phi$ is the latitude,  $\arccos(\cos \theta \cos \phi)$ is the viewing angle 
and the term $(1-\mu(1-\cos \phi \cos \phi)$ accounts for limb darkening. For the calculations 
in this paper a limb darkening coefficient of $\mu =0.6$ is used (Claret, Diaz-Cordoves\& Gimenez
  1995), but the results are insensitive to this parameter.}

In this model the number of cells in each latitudinal strip is
rounded to the nearest integer.  Hence the solid angle of the cells is
not exactly $\lambda^2$ and the number of cells is not exactly
$4\pi/\lambda^2$. The fractional error in the average cell solid angle
introduced by this approximation varies with scale length from $\simeq
0.5$\,percent for $\lambda = 0.01$\,radians to $\simeq 2$\,percent for
$\lambda = 0.5$\,radians. This level of error is negligible in the
context of this paper when compared to the much larger uncertainties in
the fits to measured data.

\begin{figure}
\centering
	\includegraphics[width = 60mm]{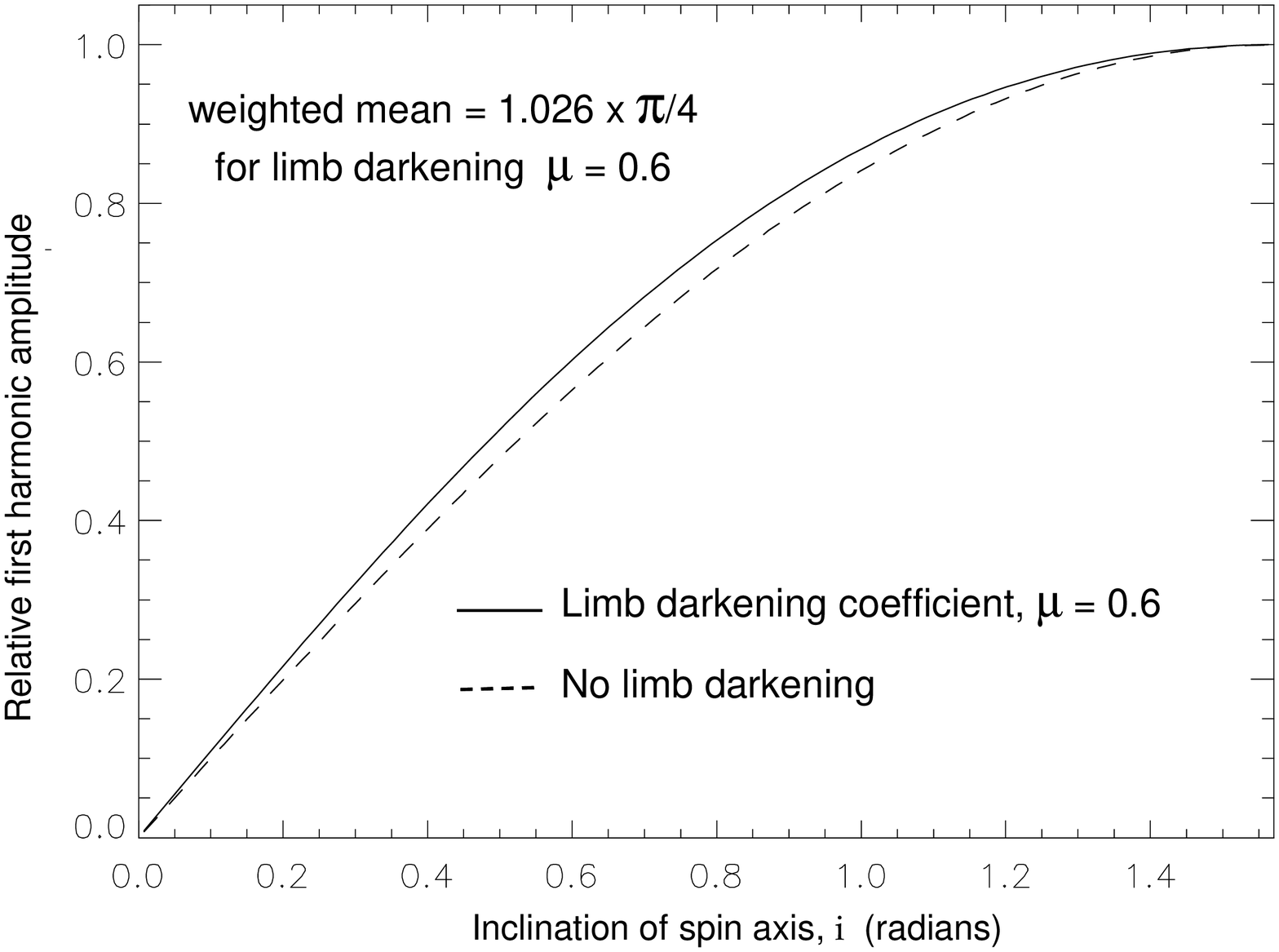}
	\caption{Variation of the light curve modulation amplitude with inclination of the stellar spin axis.}
	\label{fig2}
\end{figure}

The calcuation procedure above is valid for a star at inclination $i=90$\,deg. 
The modulation amplitude for randomly spotted stars viewed at other
inclinations is attenuated  by a factor  
that approximates to $\sin i$.  
To calculate the attenuation factor taking account of limb darkening it is
sufficient to evaluate the attenuation factor for a single spot as a
function of inclination and spot latitude which can then be averaged
over the stellar surface. Consider a spot at coordinates ($\theta$,
$\phi$) on the surface of a star of unit radius and spin axis 
inclination $i$. Viewed in Cartesian
co-ordinates, with $z$ measured along the line towards  the
observer, a point $x,y,z$ is visible if $z>0$, where
\begin{eqnarray}
	x &=&  \sin i \sin \theta - \cos i \cos \theta \cos \phi \, ,\nonumber \\
	y &=& \cos \theta \sin \phi\, , \nonumber \\
	z &=& \cos i \sin \theta  + \sin i \cos \theta \cos \phi\, .
\end{eqnarray}
\noindent{Taking account of limb darkening, the relative luminosity of the spot is given by  
$\cos q (1- \mu(1-\cos q))$ where $q$ is the angle of the normal at point ($x,y$) to the line 
of sight and hence the relative contribution of a unit area starspot to the light curve amplitude is given by
\begin{equation}
\rho _i (\theta ) = 2\int \cos q (1- \mu(1-\cos q)) \cos \phi \;
d\phi\, ,
\end{equation}
where $q=\arcsin(\sqrt{x^2+y^2})$ and $\cos\phi > -\tan\theta / \tan i$
(i.e. $z>0$).  In our model, the spots are uniformly distributed over
the stellar surface, hence their probability density varies as $\cos
\theta$, giving a weighted average of
\begin{equation}
	\overline{\rho} _i = 2\int \rho_i(\theta ) \cos \theta \; d\theta
\end{equation}

\noindent{Figure 2 shows a plot of $\overline{\rho} _i /
\overline{\rho} _{\pi /2}$ as a function of inclination. For randomly
distributed spots and $\mu = 0.6$, this
angular function is within a few per cent of $\sin i$, with a weighted mean of
$1.026\pi/4$. To a good approximation, the light curve amplitudes of
stars viewed at right angles to the line of sight can be scaled by this
mean value to give the distribution of light curve amplitudes averaged
over all viewing angles.

An implicit assumption here is that the spin axes of
  the stars are randomly oriented in space. The validity of this
  assumption was discussed in detail by Jackson \& Jeffries (2010a) for
  collections of stars, similar to those discussed here, in the young
  Pleiades and Alpha Persei clusters. Whilst the assumption of
  randomness is difficult to confirm, there is certainly no evidence
  for any strong intrinsic alignment. As we are considering a complete
  sample of stars in NGC~2516, whether they exhibit 
  periodic modulation or not, we do
  not expect any observational selection bias in the inclination angles either (see
  Jackson \& Jeffries 2012). The
  effect of any alignment would be to alter the predicted light curve amplitudes
  by a factor given by the y-axis values in
  Fig.~2 divided by the average value of $1.026\pi/4$. In the
  absence of any evidence for spin axis alignment in NGC~2516 or any
  other cluster, we do not consider this further.

Finally, the model light curve amplitude distribution is multiplied by
the previously defined completeness function, which is the probability
that a period will be measured for a given light curve amplitude (see
Fig. 3). This gives the probability density of \textit{measured} light
curve amplitudes normalised to the total number targets. 
It is this latter probability density that can be compared
directly with observational data.

\begin{figure}
\centering
	\includegraphics[width = 65mm]{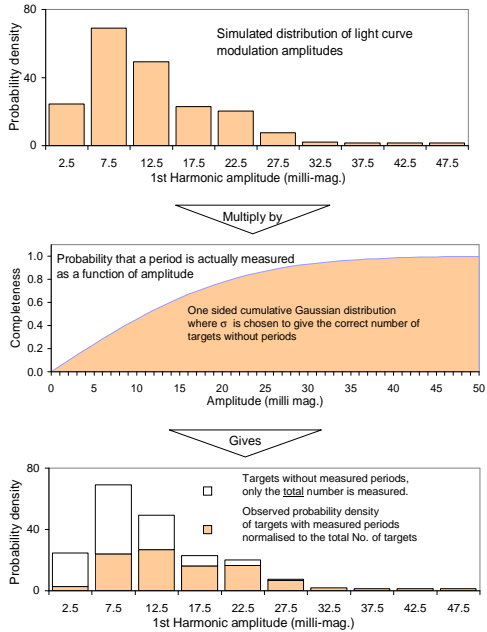}
	\caption{The process used to scale the simulated values of the probability density of light curve amplitudes to give a probability density function for stars with measured periods that can be compared with observations.}
	\label{fig3}
\end{figure}

\subsection{Variation of light curve modulation amplitudes with scale length and filling factor.}
Figure 4 summarises the properties of model light curve amplitude
distributions predicted for sets of stars with a random distribution of
starspots. These are the full amplitude distributions before scaling
by any completeness function.  The left hand plot shows the predicted
probability density of modulation amplitudes for stars with
$\gamma= 0.3$, $\kappa=0.16$  ($T_s/T_o=0.7$ for uniform spots) 
and for $\lambda = 2, 4$ and 8 degrees.
The right hand plot shows the variation of mean amplitude for a wide
range of $\lambda$ and $\gamma$.  These results indicate that to
achieve the small mean amplitudes characteristic of the active low mass stars
considered here (typically 0.015 mag, see section 3), together with high spot filling factors, requires
small values of scale length. For example a filling factor of $\gamma =
0.3$ would correspond to $\lambda \simeq 3$ degrees.  For these small scale
lengths the mean amplitude,  for a fixed value of $\kappa$, scales roughly as $\lambda \gamma^{1/2}$.

\section{Constraining starspot size using measured light curve amplitudes}
In this section, model results are compared to measured
light curve amplitude distributions for a sample of active low mass
stars in NGC 2516. In principle we
would like to constrain the four parameters, scale length, $\lambda$,
filling factor $\gamma$, luminosity ratio, $\kappa$ and completeness
scale $\sigma$. It turns out that $\lambda$ and $\sigma$ can be
determined from  fitting the model to 
the measured amplitude distribution but the
results depend to some extent on $\gamma$ and $\kappa$, which must be
estimated from other observations of the target population or more
general considerations.

\subsection{Measured distributions of light curve amplitudes}
NGC~2516 is a young (150~Myr) open cluster with a population of low
mass stars ($0.2 \leq M/M_{\odot} \leq 0.7$) 
approaching  or on the zero age main sequence. $I$-band light
curve amplitudes and rotational periods were measured for a large
sample of candidate members by Irwin et al. (2007).
A spectroscopic survey was used by Jackson \& Jeffries (2010b) and 
Jackson \& Jeffries (2012) to confirm membership for 210 stars with rotation
periods and 144 stars where no period was found. In these papers it was
shown that there were no significant differences in the
colour-magnitude diagrams, the projected equatorial velocity
distributions or the levels of chromospheric 
magnetic activity for these two subsets.

Figure 5 shows the first harmonic light curve amplitude distributions
(for the stars with measured periods) and summarises the mass range,
fraction of stars with a measured period and the mean first harmonic
light curve amplitudes (all taken from Jackson \& Jeffries 2012) for
the low-mass stars in this sample ($M_I > 7.3$, $M<0.57\,M_{\odot}$).
All these stars, and also the stars without measured periods in this
magnitude range, show saturated levels of chromospheric activity.  
  This is important, because it means we do not expect the filling
  factor of spots to vary with rotation rate and can treat the sample
  as a single population. About 40 per cent of the sample have
  $M<0.35\,M_{\odot}$ and may be fully convective, according to a
  theoretical mass-magnitude relationship from Baraffe et al. (2002).
The results are shown in four equal bins of absolute I magnitude. The
proportion of stars with and without measured periods in each bin are
true estimates, corrected for any bias due to the preferential
targeting of stars with measured periods in the spectroscopic sample
(see Table 5 of Jackson \& Jeffries 2012). The form of the amplitude
distributions mirrors the model distributions shown in Fig. 3, with an
initial increase in frequency (moderated by the completeness
function described in section 2), a peak in the range 0.01 to 0.02 mag followed by a rapid
decay, with no measured amplitudes $>0.05$ mag. The error bars on the
measured data represent Poissonian uncertainties.
   
\begin{figure}
\centering
	\includegraphics[width = 87mm]{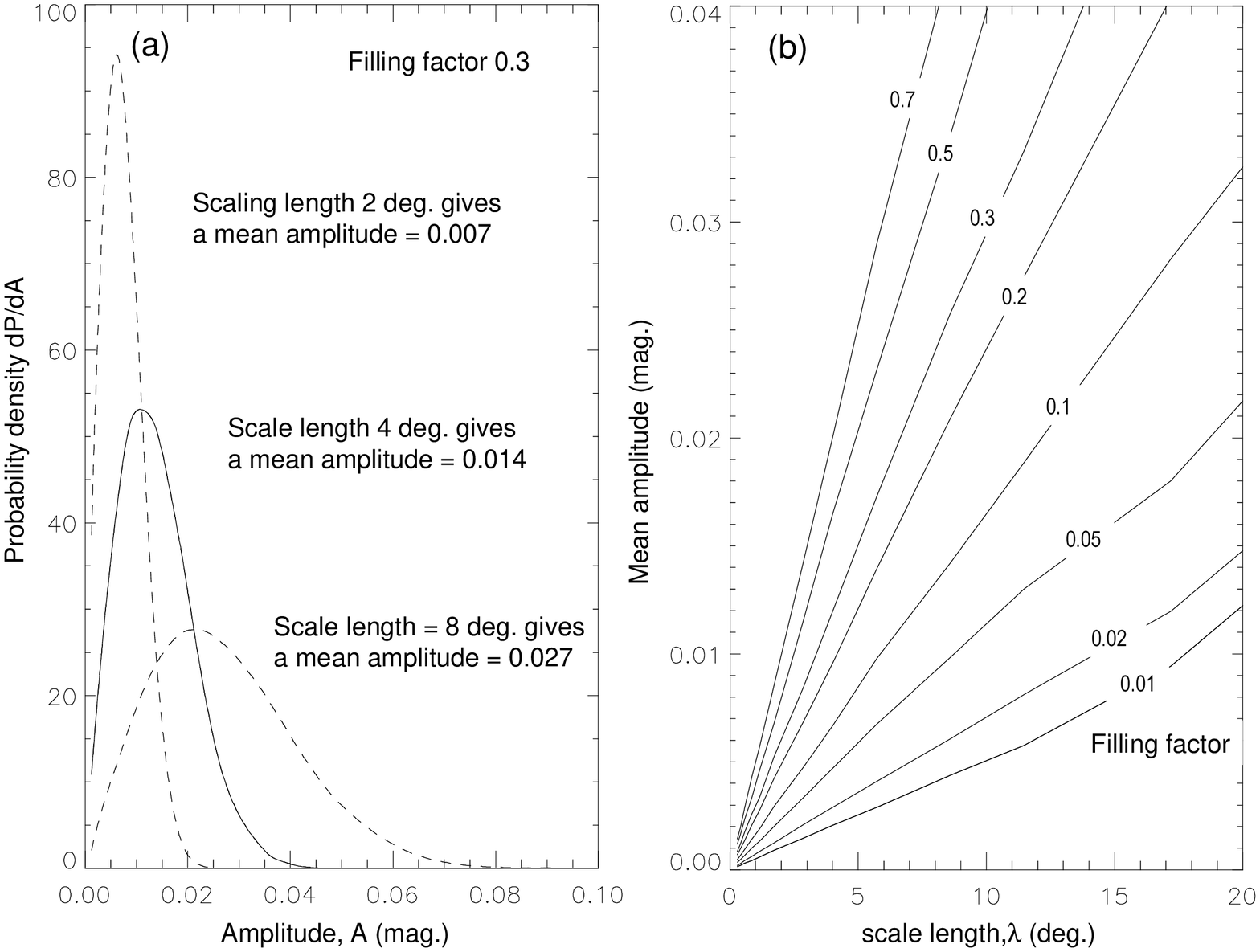}
	\caption{Monte Carlo simulations of light curve amplitudes for
          stars with a large number of randomly distributed starspots
          of uniform area where the spotted areas have a luminosity ratio 
          of $\kappa=0.16$~relative to the
          unspotted photosphere. Plot (a) shows the probability density
          for stars with a filling factor $\gamma=0.3$ for various spot
          scale lengths. Plot (b) shows the mean light curve amplitude 
          as a function of scale length for various filling factors.}
	\label{fig4}
\end{figure}

\begin{figure}
\centering
	\includegraphics[width = 80mm]{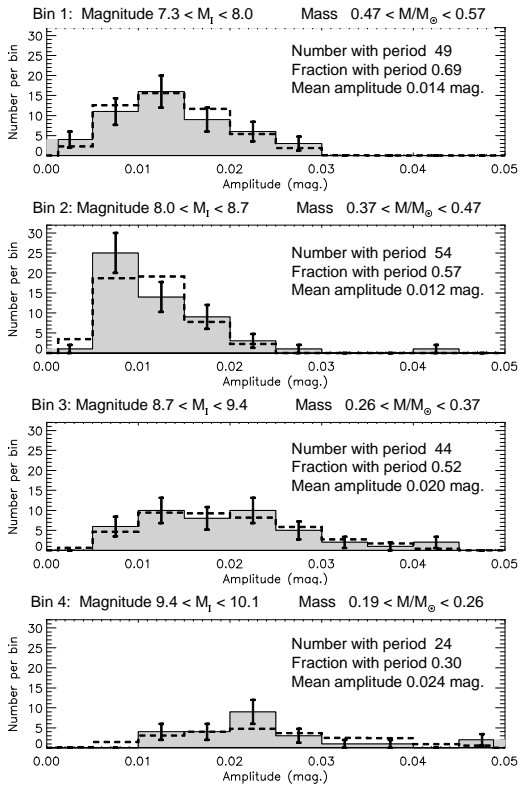}
	\caption{The probability density of light curve modulation
          amplitudes for low mass stars in  NGC~2516  (Jackson et al. 2012). Results are shown in four bins of absolute I
          magnitude with equivalent masses from the models of Baraffe
          et al. (1998 \& 2002). The shaded histogram shows the number of stars
          with measured periods as a function of first harmonic
          amplitude, together with their expected Poisson
          uncertainties. The dashed lines show numbers predicted 
          from a Monte Carlo simulation of a randomly spotted star
          using ``best fit'' values for $\lambda$ and $\sigma$ (see
          section 3.2), assuming $\gamma=0.3$ and $\kappa=0.16$.}
	\label{fig5}
\end{figure}

\subsection{Constraining scale length and completeness scale}
A maximum likelihood technique is used to constrain possible values of
$\lambda$ and completeness scale $\sigma$, as a function
of $\gamma$ and $\kappa$. To this end
a grid of probability densities for the light curve amplitudes 
is generated in 0.005 magnitude bins
from 0 to 0.1 mag for $\lambda = 1$~to $100\sqrt \gamma$~degrees and $\sigma = 0$~to $0.1$~for a series
of values of $\gamma$ and $\kappa$ (see below). As there are less than 20 stars in
  each bin, chi-squared methods would yield biased results, so
the most probable fit is found using the
modified form of the Cash statistic (Cash 1979). 
\begin{equation}
 C = 2\sum^N_{i=1}\left[y(x_i) - y_i + y_i(\ln y_i -\ln
   y(x_i))\right]\, 
\end{equation}
where $y_i$ are the observed data and $y(x_i)$ are the model values.
This form of the Cash statistic  is appropriate for binned data
  with small counts and
can be treated in the same way as $\chi
^2$ to determine goodness of fit and
determine the confidence levels of fitted parameters.

Figure 6 shows contour plots of the modified Cash statistic constructed
by modelling each of
the data sets in Fig. 5. The vertical axis shows the completeness
scale, $\sigma$ and the horizontal axis is the scale
length, $\lambda$.  Contours mark 95 per cent confidence limits
around the the combination of $\sigma$ and $\lambda$ that
best fit the observations. Results are shown for four possible
values of filling factor, $\gamma = 0.05, 0.1, 0.3, 0.5$ at a fixed luminosity
ratio of  $\kappa = 0.16$ (equivalent to $T_s/T_0 = 0.7$ for uniform
spots).  Notes on each plot show the minimum value of
the Cash statistic for each case 
and the average value of the resultant probability that the best fit
model is a good fit to the measured distribution. 
Examples of the model predictions are shown in Fig.~5 for the case of
$\gamma=0.3$ and $\kappa=0.16$. It is important to note (see
the $C_{\rm min}$ values in Fig.~6) that all the combinations of
$\gamma$ and $\kappa$ that we tested yield statistically acceptable
fits with an appropriate choice of $\lambda$ and $\sigma$. That is,
these observations alone are incapable of constraining $\gamma$ or
$\kappa$.

\begin{figure*}
\centering
\begin{minipage}[htb]{0.9\textwidth}
	\includegraphics[width = 145mm]{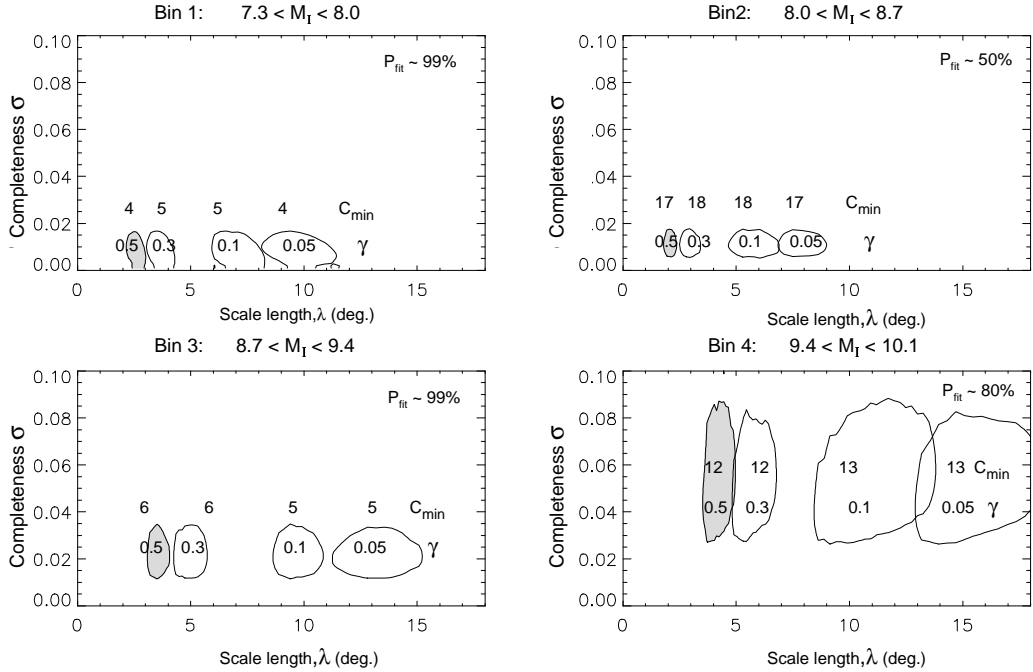}
\end{minipage}
	\caption{Contour plots of the Cash statistic as a function of
          scale length, $\lambda$ and completeness function
          $\sigma$. Contours indicate the 95 per cent
          interval around the combination of $\lambda$ and $\sigma$
          that best fits the four subsets of measured data in Fig.~5.
          Results are shown for a luminosity ratio $\kappa=0.16$ at four values of filling factor, $\gamma$. Also shown are the                   minimum value of the Cash statistic for each case and the average probability of fit to each subset of measured data.}
	\label{fig6}
\end{figure*}

Figure 6 shows that the value of $\sigma$ is reasonably independent of $\gamma$.
However, $\sigma$ does vary, as expected, 
with absolute magnitude, increasing from 0.01 to 0.025~mag
over the first three subsets and rising sharply to 0.06~mag for the
faintest subset.  This corresponds to completeness values of 0.97, 0.93,
0.66 and 0.33 for light curves of amplitude 0.02 mag.  This is reasonably
consistent with the expected variation in measured period completeness with
magnitude for stars in the Irwin et al. (2007) survey from which the
data were taken.

The most likely value of $\lambda$ varies as
$\simeq \gamma^{-1/2}$ and also perhaps weakly with absolute $I$
magnitude. Figure 7 shows a more detailed plot of the variation of the
best fitting $\lambda$ as $\gamma$ is varied. The uncertainties shown on this plot
correspond to 68 per cent confidence intervals in one parameter.

Finally, Fig. 8 summarises the results obtained for different values
of $\kappa$ ($0.03\leq\kappa\leq0.59$, corresponding to $0.5\leq
T_s/T_0<0.9$ for uniform spots). The error bars here
indicate the largest and smallest values of $\lambda$ obtained from the
four $M_I$ subsamples in Fig.~\ref{fig5}. This plot shows that for a given
filling factor, the best-fitting value of $\lambda$ increases with
$\kappa$  and is roughly proportional to $(1-\kappa)^{-1}$.

\begin{figure}
\centering
	\includegraphics[width = 80mm]{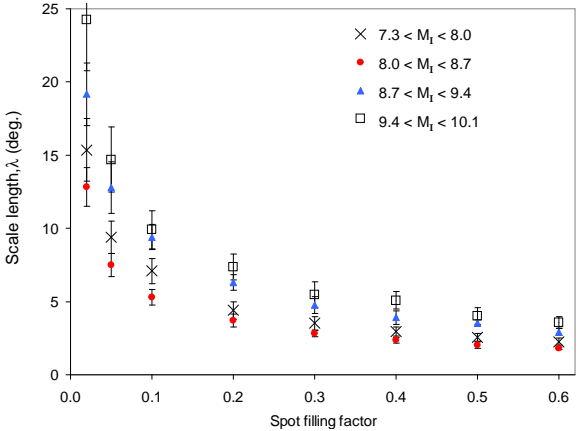}
	\caption{The variation of the best-fitting 
          scale length as a function of
          with spot filling factor, for a spot luminosity ratio of
          $\kappa=0.16$. The results are shown for the data in four
          subsamples of
          absolute I-magnitude (see Fig.~\ref{fig5}). The error bars indicate 68
          per cent confidence intervals.}
	\label{fig7}
\end{figure}

\begin{figure}
\centering
	\includegraphics[width = 80mm]{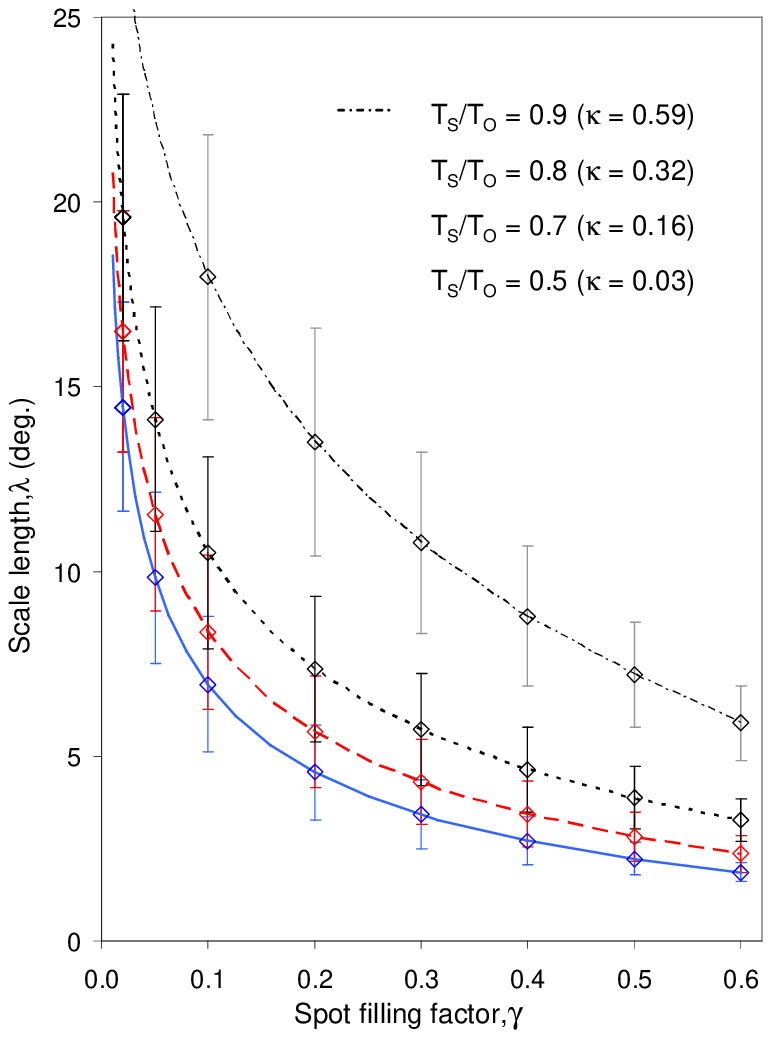}
	\caption{Possible constraints on the scale length for
          independent areas of starspot activity as a function of
          filling factor, $\gamma$ and temperature ratio
          $T_s/T_o$. Lines indicate the mean value of $\lambda$ that
          best fits measured data shown in Fig.~5. The error bars
          indicate the highest and lowest $\lambda$ values obtained
          from fitting the data in each of the four absolute I-magnitude
          subsamples in Fig.~\ref{fig5}.}
	\label{fig8}
\end{figure}

\section{Discussion}

The main motivation for these simulations was the suggestion by Jackson \&
Jeffries (2012) that the small light curve amplitudes seen in the young, active
low-mass stars of NGC~2516 were compatible with large spot
coverage fractions, and might constrain the typical spot size. The
results shown in Fig.~\ref{fig8} confirm this idea. Both the small
observed light curve amplitudes of the periodic stars 
and the fraction of cluster members for which
rotation periods could not be found can be explained by a
random distribution of small, dark spots on the stellar surface.
Jackson et al. (2009) estimated that 50 per cent or more surface coverage by
dark spots ($T_s/T_o=0.7$) may be required in the coolest stars  of
  this sample} to explain their
large radii compared with standard evolutionary models. Fig.~\ref{fig8} 
shows that such spot coverage leads to inferred spot scale lengths of $3 \pm 1$
degrees.

The estimated characteristic spot scale length depends on (a) the
assumed spot coverage fraction and spot temperature and (b) the assumption of a
random spot distribution.  Neither of these can be independently
constrained by just the light curve data. Fig.~\ref{fig8}
demonstrates that the best-fitting spot scale length could be
significantly larger if the spot coverage were smaller or 
if the spots were hotter.  The light curve amplitude
distributions alone  could equally well be explained by $\sim 20$ randomly
placed spots of diameter 10 degrees ($\gamma = 0.05$) or $\sim 2500$
spots of diameter 3 degrees ($\gamma = 0.5$).

The indirect estimate of spot coverage and temperature from Jackson et
al. (2009) assumed that the inflated radii observed for these
magnetically active stars are solely due to starspots. This followed
empirical evidence that larger radii are also seen in the low-mass
components of fast-rotating, eclipsing binary stars and that this
radius discrepancy has been linked with magnetic activity
(L\'opez-Morales 2007). Chabrier et al. (2007) showed that 30--50 per
cent coverage by black spots could reproduce these results
too. However, other effects, such as a reduction of convective
efficiency or inhibition of the onset of convection by interior
magnetic fields might also increase radii and thus reduce the required
spot coverage (see also MacDonald \& Mullan 2012). Hence these indirect
estimates of $\gamma$ are possibly upper limits.

Independent determinations of the spot coverage in active stars depend
crucially on the technique used. Analyses of photometric light curves
or Doppler imaging maps probably underestimate total spot coverage
because of their limited spatial resolution or lack of sensitivity to
axisymmetric spot distributions (see Solanki \& Unruh 2004). The most
direct estimates of $\gamma$ and $T_s/T_0$ are likely to come from
measuring a number of TiO absorption bands in high
resolution spectra and fitting them with two-temperature models, using
the spectra of magnetically inactive stars as templates (see O'Neal et
al. 1998). Results are reported for a number of very active G-
and K-type stars by O'Neal et al. (2004) and O'Neal (2006). These
include three active young dwarf stars, EK~Dra (G1.5V), LQ~Hya (K0V) and
EQ~Vir (K5V), for which $\gamma = 0.4\pm 0.1$ and $T_s/T_0 = 0.70 \pm
0.05$ were determined.  Unfortunately, the same technique is
ineffective for M-dwarfs since their unspotted photospheres also show
strong TiO absorption bands (O'Neal et al. 2005).

If we were
to extrapolate and assume that similar parameters ($\gamma= 0.4 \pm
0.1$, $T_{s}/T_0 = 0.70\pm 0.05$) apply to the active M-dwarfs of
NGC~2516, then a scale length $\lambda = 3.5^{+2}_{-1}$ degrees is implied
(see Fig.~\ref{fig8}). The legitimacy of this extrapolation could be
questioned on the basis that the stars studied by O'Neal et al. were of
earlier spectral type, with shallower convection zones. As discussed in
section~2, it is possible that the nature of the dynamo changes as
the convection zone deepens and especially when stars become fully
convective, which may be the case for the coolest 40 per cent of stars (roughly
the third and fourth bins in Fig.~5) in our
sample. However, the vast majority of our sample are M0--M4 dwarfs for
which there is reasonable evidence that magnetic activity is generated
and manifested in a similar way to K-dwarfs. This includes the similarity of 
rotation-activity relationships
between G-, K- and M-dwarfs as cool as type M4 (Jeffries et al. 2011;
Reiners, Joshi \& Goldman 2012) 
and the similar spot filling factors and distributions in active K- and early M-dwarfs
inferred from Doppler tomography (Barnes \& Collier Cameron 2001 and
see below). Counter to this, there is some evidence from Zeeman Doppler
imaging that the large scale magnetic field does undergo a change
towards a more axisymmetric, poloidal topology at the fully convective
boundary (beyond type M3; Morin et al. 2008, 2010). How this relates to photospheric
fields at the scale of starspots is
unknown, although the bulk of magnetic energy still appears to reside at
smaller size scales (Reiners \& Basri 2009). In summary, it is possible
our extrapolation is invalid for the lowest mass stars of our sample.

The assumption of random spot coverage is the simplest approach we
could have adopted, but as discussed in 
section~2, there is some evidence that real low-mass
stars may behave differently. Doppler images have revealed long-lived
spots, or unresolved spot groups, at high latitudes or covering the
rotational poles for some young K-dwarf stars at some epochs
(e.g. on the rapidly rotating K-type ZAMS stars AB Dor in 1993/1994,
Jeffers et al. 2007 and BO~Mic in 2002, Barnes 2005), 
but not others (e.g. AB~Dor in 1989, K\"urster, Schmitt
\& Cutispoto 1994; BO~Mic in 1998, Barnes et al. 2001).
A concentration of spots towards high latitudes
would reduce light curve amplitudes for a given $\gamma$, so larger
spot scale lengths would be required to compensate. However, the effect
would not be large; even if half the spot coverage were concentrated in
an axisymmetric polar cap this would only increase the required $\lambda$
by a factor of $\sqrt{2}$. In any case, the situation for fast-rotating
M-dwarfs may be different. Doppler images of HK~Aqr and EY~Dra, M1-2
dwarfs with rotation periods $<1$ day, reveal spots either
at low latitudes or with no clear latitude dependence at all (Barnes \&
Collier Cameron 2001). Longitudinal asymmetries or preferential spot
longitudes are more difficult to assess. For example if the presence of
a spot or spot group at one longitude made it more likely that
further spots would emerge at similar longitudes then this would
increase photometric modulation for a given $\gamma$ and alter the
relationship between $\lambda$ and $\gamma$ in Fig.~\ref{fig8}. Any
attempt to observationally identify ``active longitudes'' in single
stars is hampered by the possibility of differential rotation and a
lack of spatial resolution. 

The spot scale length implied by our simple model can be converted to a
linear scale length if the stellar radius is known. Jackson et al
(2009) estimated radii of 0.4--0.6\,$R_{\odot}$ for the stars in
NGC~2516 considered here, which for $\lambda \simeq3.5$ degrees implies
absolute scale lengths of order 25\,000\,km.  Baumann \& Solanki (2005)
have studied the distribution of spot sizes on the Sun, finding that
both the sizes of individual spots and of spot groups are
well-represented by log-normal distributions. A starspot area of
$6\times 10^{14}$\,m$^2$ is a factor of 5--6 larger than the modal area
(umbra plus penumbra) of individual sunspots ($\simeq 1$ degree
diameter), but only a factor of 2--3 larger than a typical sunspot
group and well within the observed dispersion.

High cadence, high signal-to-noise ratio photometry is now capable of
estimating the sizes of individual starspots or starspot groups in
systems where the spots are occulted by an exoplanet. Wolter et
al. (2009) and Silva-Valio et al. (2010, 2011) have analysed modulation of
the light curve during exoplanetary transits of a reasonably rapidly
rotating (P=4.5~days) active G7V host, CoRoT-2.  They found that
typically the exoplanet transits 5 spots as it crosses the stellar disc
and that these spots (or spot groups) have a  diameter of  40\,000 to 150\,000\,km 
with temperatures in the range 3600 -- 5000\,K (for $T_{eff} = 5625K$) giving
$0.6 < T_s/T_o < 0.9$} and that the transited (low latitude)
stellar region has a 10--20 per cent spot coverage.  At these relatively
low filling factors, the ``spot size'' is a roughly equivalent
parameter to the scale length considered in this paper.
Bearing in mind that the quality of the CoRoT data limited the analysis to spot sizes
larger than 30\,000\,km  and since there is some degeneracy between spot
size and spot temperature in both their and our analyses, it seems that
these spots may be only a little larger than those we have deduced for
the very active M-dwarfs in NGC 2516. Alternatively, we could reverse
this chain of argument and say that if the spots on the NGC~2516 stars
were of a similar absolute size and temperature  ratio to those of CoRoT-2, and randomly
distributed over the stellar surface then from Fig.~8, the spot filling factor
would be $\gamma < 0.1$. Alternatively, for a spot coverage of $\gamma \simeq0.4$
 then $T_s/T_o$ would need to be $ > 0.9$. 

In summary this paper shows that both the small light curve amplitudes
observed in a set of fast-rotating, young, magnetically active
M-dwarfs, and the lack of rotational modulation seen in a large
fraction of their siblings, {\it could} be explained by a starspot
model consisting of large filling factors of dark spots that are
randomly distributed on the stellar surface. If the M-dwarfs in NGC~2516 have a spot coverage
fraction $\gamma \sim 0.4\pm0.1$ and a spot/photosphere temperature ratio of
$T_s/T_0 \sim 0.7\pm0.05$, as suggested by extrapolation of the TiO
modelling of very active K dwarfs (O'Neal  et al. 2004, 2006),
then the scale length between independent areas of
starspot activity is $\lambda \simeq 3.5^{+2}_{-1}$\,degrees (or 25\,000\,km). This
scale length varies as $\gamma^{-1/2}$ and increases with
the assumed spot temperature, neither of which can be constrained by
the light curve data. There is an urgent need to independently
determine these parameters in lower-mass active stars, both to address
the issue of typical spot sizes and also to assess the possible influence of
spots in inflating stellar radii above the predictions of current
evolutionary models.  The spot scale lengths found above are only a
little larger than typical sunspot groups but a little smaller than the
spot sizes so far inferred from mapping using transiting exoplanets.  
If these small spot scale lengths are confirmed then this complicates
 the interpretation of Doppler and Zeeman Doppler imaging maps, 
 where typical angular resolutions are in the range 3 -- 10 degrees.

\section{acknowledgements}
Based on observations collected at the European Southern
Observatory,Paranal, Chile through observing programs 380.D-0479 and
266.D-5655. RJJ acknowledges receipt of a Wingate scholarship.

\nocite{Baraffe1998a}
\nocite{Baraffe2002a}
\nocite{Barnes2001a}
\nocite{Barnes2001b}
\nocite{Barnes2005b}
\nocite{Barnes2011a}
\nocite{Baumann2005a}
\nocite{Berdyugina2005a}
\nocite{Cash1979a}
\nocite{Chabrier2007a}
\nocite{Chabrier2006a}
\nocite{Claret1995a}
\nocite{Irwin2007a}
\nocite{Jackson2009a}
\nocite{Jackson2010a}
\nocite{Jackson2010b}
\nocite{Jackson2012a}
\nocite{Jeffers2006a}
\nocite{Jeffers2007a}
\nocite{Jeffries2011a}
\nocite{Kuerster1994}
\nocite{Lockwood2007a}
\nocite{LopezMorales2007a}
\nocite{Macdonald2012a}
\nocite{Morales2009a}
\nocite{Morales2010a}
\nocite{Morin2008a}
\nocite{Morin2010}
\nocite{Oneal1998a}
\nocite{Oneal2004a}
\nocite{Oneal2005a}
\nocite{Oneal2006a}
\nocite{Parker1975a}
\nocite{Radick1998a}
\nocite{Rockenfeller2006a}
\nocite{Ribas2008a}
\nocite{Schuessler1992a}
\nocite{Silva2010a}
\nocite{Silva2011a}
\nocite{Solanki2004a}
\nocite{Spruit1986a}
\nocite{Stout-Batalha1999}
\nocite{Strassmeier2009a}
\nocite{Thomas2008a}
\nocite{Torres2010a}
\nocite{Wolter2009a}
\nocite{Reiners2009b}
\nocite{Reiners2012a}
\nocite{Reiners2010b}

\bibliographystyle{mn2e} 
\bibliography{references}


\bsp 

\label{lastpage}

\end{document}